\documentclass{ws-p8-50x6-00}

\begin{document}

\title{Vector Condensation at Large Chemical Potential}

\author{F. Sannino and W. Sch\"afer}

\address{NORDITA, Blegdamsvej 17, DK-2100 Copenhagen \O, Denmark\\
E-mail: francesco.sannino@nbi.dk, schaefer@nordita.dk}


\maketitle

\abstracts{ We discuss the condensation of relativistic spin-one
fields at high chemical potential. This phenomenon leads to the
spontaneous breaking of rotational invariance, together with the
breaking of internal symmetries.}

\section{Introduction}

There has been much interest recently in the phase structure of
Quantum Chromodynamics (QCD) and QCD--like theories. An overview
has been given at this workshop \cite{FSGiselda}. Here the phase
diagram not only depends on temperature and density, but also on
the number of flavors $N_f$, and colors $N_c$. Most of our
knowledge on the QCD phase diagram at finite temperature derives
from lattice simulations \cite{Hands:2001jn}. Lattice studies of
the high density, low temperature region are however seriously
obstructed by a complex fermion determinant . No such difficulties
exist for $N_c = 2$, and hence predictions for two--color QCD can
be compared with numerical solutions from the lattice. Such a
theory with $N_f$ flavors has a global $SU(2N_f)$ symmetry. There
are color--singlet diquark states which play the role of baryons.
Interestingly, it has been predicted\cite{Lenaghan:2001sd} that at
sufficiently high baryon chemical potential, spin-1 diquark states
will condense. Indeed there appears to be preliminary evidence for
this phenomenon  from lattice calculations \cite{MPLGiselda}. With
this physical application in mind we shall study a vector field
transforming in the adjoint representation of a global $SU(2)$
group \cite{SS02}. The chemical potential is chosen along one of
the generator's direction (the ``baryon charge" operator), and
breaks the global symmetry explicitly to $U(1) \times Z_2$.
Apparently condensation of a vector field singles out a direction
in space, the rotational $SO(3)$--invariance is spontaneously
broken to an $SO(2)$ subgroup.

Introduction of a chemical potential explicitly breaks the Lorentz
invariance, and in such a situation the relation between broken
symmetries and Goldstone modes (gapless excitations) is of special
interest. Here the particle--physics textbook folklore that every
broken generator corresponds to a Goldstone boson does not apply.
Instead, following Nielsen and Chadha, one should properly
distinguish between two types of Goldstone modes, depending on
whether in the long--wavelength limit their energy varies with an
odd (type I) or even (type II) power of momentum. The following
inequality holds:
\begin{equation}
N_{bg} \leq N_I + 2 \cdot N_{II} \, , \label{NC}
\end{equation}
between the numbers of broken generators $N_{bg}$, and type I/II
Goldstone modes $N_{I/II}$ \cite{NC76}. In order to elucidate
these aspects in case of the vector condensation, we proceed with
constructing the simplest effective Lagrangian for a vector field.

\section{A lagrangian model for vector condensation}

We introduce a vector field $A_\nu^a, a=1,2,3$ of mass $m$
transforming under the adjoint representation of a global $SU(2)$
symmetry. The Lagrangian, endowed with a potential of fourth order
in the field, reads:
\begin{eqnarray}
{\cal
L}&=&-\frac{1}{4}F^a_{\mu\nu}F^{a{\mu\nu}}+\frac{m^2}{2}A_{\mu}^a
A^{a\mu} -
 \frac{\lambda}{4}\left(A^a_{\mu}A^{a{\mu}}\right)^2 +
 \frac{\lambda^{\prime}}{4} \left(A^a_{\mu}A^{a{\nu}}\right)^2 \ ,
 \end{eqnarray}
with $F_{\mu
\nu}^a=\partial_{\mu}A^a_{\nu}-\partial_{\nu}A^a_{\mu}$, and
metric convention $\eta^{\mu \nu}={\rm diag}(+,-,-,-)$. Here we
confine ourselves to positive $\lambda$ and $\lambda^{\prime}$.
Stability of the potential furthermore demands $\lambda >
\lambda'$. The effect of a nonzero chemical potential associated
to a given conserved charge - related to the generator $B=T_3$ -
can be readily included \cite{Lenaghan:2001sd} by modifying the
derivatives acting on the vector fields:
\begin{equation}
\partial_{\nu} A_{\rho} \rightarrow \partial_{\nu}A_{\rho} - i
\left[B_{\nu}\ ,A_{\rho}\right]\ ,
\end{equation}
with $B_{\nu}=\mu \,\delta_{\nu 0} B\equiv V_{\nu} B$ where
$V=(\mu\ ,\vec{0})$. The chemical potential thus resembles an
external field in temporal direction.
The kinetic term can then be cast in the form
\begin{eqnarray} {\cal
L}_{kinetic}&=&\frac{1}{2}A_{\rho}^a \, {\cal{K}}_{ab}^{\rho\nu}
\, A_{\nu}^b
\end{eqnarray}
with
\begin{eqnarray}
{\cal{K}}_{ab}^{\rho\nu} &=& \delta_{ab} \left[g^{\rho\nu}
\partial^2 -\partial^{\rho}\partial^{\nu}\right] -
4i\gamma_{ab}\left[g^{\rho \nu}V\cdot \partial - \frac{V^{\rho}
\partial^{\nu} + V^{\nu} \partial^{\rho} }{2} \right] \nonumber \\ &+& 2
 \chi_{ab}\left[V\cdot V g^{\rho\nu}-V^{\rho}V^{\nu}\right] \, ,
\label{kinetic}
\end{eqnarray}
where
\begin{equation}
\gamma_{ab}={\rm Tr}\left[T^a\left[B,T^b\right] \right]\ , \qquad
\chi_{ab}={\rm
Tr}\left[\left[B,T^a\right]\left[B,T^b\right]\right] \ .
\end{equation}
{}For $B=T^3$ we have \begin{equation}
\gamma_{ab}=-\frac{i}{2}\epsilon^{ab3} \ , \quad
\chi_{11}=\chi_{22}=-\frac{1}{2} \ , \quad \chi_{33}=0 \ .
\end{equation} The chemical potential induces a ``magnetic-type''
mass term for the vectors at tree-level. The symmetries of the
potential are more easily understood using the following Euclidean
notation:
\begin{equation}
\varphi_M^a=(A_M^1,A_M^2) \ , \qquad  \psi_M=A_M^3 \ ,
\end{equation}
with $A_M=(iA_0,\vec{A})$ and metric signature ($+,+,+,+$).  In
these variables the potential reads:
\begin{eqnarray}
V_{Vector}&=&\frac{m^2}{2}\left[|\vec{\varphi}_{0}|^2  + \psi_M^2
\right] +\frac{m^2-\mu^2}{2} |\vec{\varphi}_I|^2
+\frac{\lambda}{4}\left[|\vec{\varphi}_{M}|^2 + \psi_M^2\right]^2
\nonumber \\&-&
\frac{\lambda^{\prime}}{4}\left[\vec{\varphi}_{M}\cdot
\vec{\varphi}_{N} + \psi_M \psi_N\right]^2
\end{eqnarray}
with $I=X,Y,Z$ while $M,N=0,X,Y,Z$ and repeated indices are summed
over. At zero chemical potential $V_{Vector}$ is invariant under
the $SO(4)$ Lorentz transformations while only the $SO(3)$
symmetry is manifest at non zero $\mu$. As noted above, the
chemical potential explicitly breaks the global $SU(2)$ symmetry
to $U(1) \times Z_2$. Note that for the special parameter choice
$\lambda'=0$, the potential alone is invariant under an $SO(6)$
rotation group.

It is apparent, that due to the presence of the term  proportional
to $\displaystyle{(m^2-\mu^2)}$, we have to distinguish the cases
$\mu \leq m$ and $\mu > m$.

\subsection{The Symmetric Phase: $0 < \mu \leq m$} Here the $SO(4)$
Lorentz and $SU(2)$ symmetries are explicitly broken to $SO(3)$
and $U(1)$ respectively by the chemical potential. All fields have
a vanishing expectation value in the vacuum:
$<\vec{\varphi}_M>=<\psi_M>=0$. The curvatures of the potential on
the vacuum (the masses) are:
\begin{eqnarray}
M^2_{\varphi_{0}^a}= M^2_{\psi_{M}}=m^2 \ , \qquad
M^2_{\varphi_{I}^a}=m^2 - \mu^2 \label{Vcurvature} \ .
\end{eqnarray}
We obtain the dispersion relations by diagonalizing the 12 by 12
matrix ${\cal{K}}$ (\ref{kinetic}). This leads to 3 physical
vectors (i.e. each of the following states has 3 components) with
the dispersion relations:
\begin{eqnarray}
E_{\varphi^{\mp}}=\pm\mu + \sqrt{\vec{p}^2+m^2} \ , \qquad
E_{\psi}&=&\sqrt{\vec{p}^2+m^2} \ .
\end{eqnarray}
We observe that for $\mu \to m$ the 3 physical components
associated with $E_{\varphi^{+}}$ will become massless signaling
an instability. Indeed, for higher values of the chemical
potential larger than $m$ a vector condensation sets in.

\begin{figure}[t]
\epsfxsize=25pc 
\epsfbox{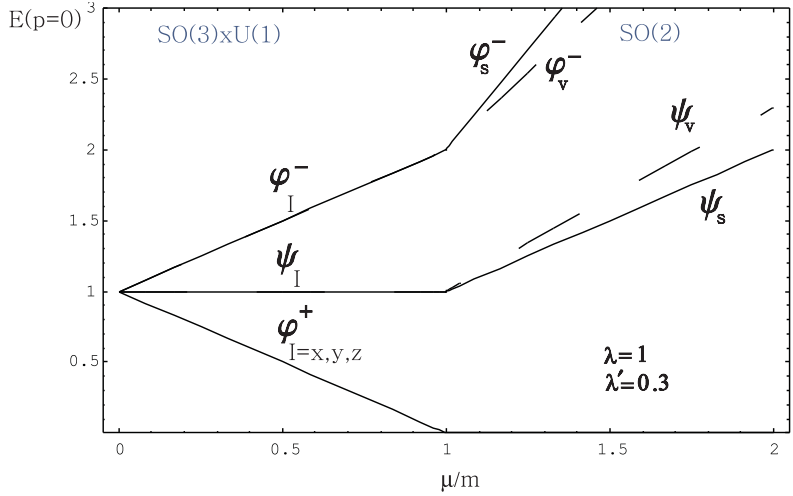} 
\caption{The mass gaps $E(0)$ as a function of the chemical
potential $\mu$. \label{fig:1}}
\end{figure}

\subsection{The Spin-Flavor Broken Phase: $\mu>m$} In this phase
at the global minimum of the potential we still have
$\displaystyle{<{\varphi}_0^a>=<\psi_M>=0}$. However there emerges
a condensate:
\begin{equation}
<{\varphi}_X^1>=\sqrt{\frac{\mu^2-m^2}{\lambda -
\lambda^{\prime}}} \ .
\end{equation}
We have a manifold of equivalent vacua which are obtained rotating
the chosen one under a $Z_2\times U(1)\times SO(3)$
transformation. The choice of the vacuum partially locks together
the Lorentz group and the internal symmetry while leaving unbroken
only the subgroup $Z_2\times SO(2)$. Two generators associated to
the Lorentz rotations are now spontaneously broken together with
the $U(1)$ generator.

To proceed with the calculation of the dispersion relation of the
vector states in the broken phase we first have to evaluate the
curvatures of the potential on the new vacuum. The explicit
formulas can be found elsewhere \cite{SS02}. We find in general
three states with null curvature. An exception occurs for the
choice $\lambda^{\prime}=0$ in which case there are five zero
curvature states. To explain this behavior we note that for
$\lambda^{\prime}=0$ the potential possesses an enhanced $SO(6)$
global symmetry which breaks to $SO(5)$ when the vector field
condenses. The associated five states would correspond to the
ordinary Goldstone modes in the absence of an explicit Lorentz
breaking. Now the symmetry breaking pattern of our theory is
$SO(3) \times U(1) \times Z_2 \to SO(2) \times Z_2$ and we find
the three expected gapless excitations irrespective of the
parameters of the potential.

In Fig.1 we show how the mass gaps evolve as a function of the
chemical potential $\mu$. In the unbroken phase, for $\mu \leq m$
all gaps are threefold degenerate and the splitting of the states
of different baryon charge is clearly visible. For $\mu > m$ the
modes with nonvanishing gap split further. The dashed lines are
twofold degenerate ($SO(2)$--vectors) while the solid lines
correspond to $SO(2)$ scalar states. There are three gapless
modes, whose dispersion relations however show some subtleties.

\subsection{Dispersion relations and Goldstone counting}

The gapless states also fall into a doublet forming an $SO(2)$
vector and an $SO(2)$ singlet. Their dispersion relations at small
momenta have the form
\begin{eqnarray}
E^2_{\varphi^+_V} = v^2_{\varphi^+_V} \, \vec{p}^2 + {\cal O}(p^4)
\, , \, E^2_{\varphi^+_S} = v^2_{\varphi^+_S} \, \vec{p}^2 + {\cal
O}(p^4) \, ,
\end{eqnarray}
where we introduced the ``superfluid velocities" $v_{\varphi^+_V},
v_{\varphi^+_S}$. One can easily show that these velocities can be
expressed directly in terms of the curvatures in the directions
orthogonal to the gapless modes as follows:
\begin{equation}
v_{\varphi^-_V}^2 = {M^2_{\varphi^2_Y} \over M^2_{\varphi^2_Y} + 4
\mu^2} \ , \qquad v_{\varphi^-_S}^2 = {M^2_{\varphi^1_X} \over
M^2_{\varphi^1_X} + 4 \mu^2} \ .
\end{equation}
It now turns out, that always $M^2_{\varphi^1_X} \neq 0$.
Therefore the scalar state will always have a linear dispersion
relation $E \propto p$. Not so the $SO(2)$ vector states. In the
case of enhanced potential symmetry $SO(6)$, the curvature $
M^2_{\varphi^2_Y}$ vanishes and with it the velocity
$v_{\varphi^-_V}^2$. The dispersion relation hence becomes a
quadratic one, $E \propto p^2$, the $SO(2)$ state turns into a
type II Goldstone boson. We emphasize, that were Lorentz
invariance unbroken, the curvatures $M_i^2$ would precisely be
masses and there would be a Goldstone for each flat direction of
the potential. Obviously the chemical potential prevents the
emergence of extra Goldstone bosons in the case $\lambda'=0$.
Simply the full Lagrangian (kinetic plus potential terms) does not
share the potential's larger symmetry. Still, the additional flat
directions of the potential have a physical effect: {\it they are
responsible for turning some of the Goldstones into type II
modes}. This appears to be a perfectly general phenomenon for the
type of kinetic term that arises at finite chemical potential.
Finally we count: for $\lambda' \neq 0$ there are three broken
generators and three type I Goldstones, hence $N_{bg}=N_I$, the
standard situation. For $\lambda'=0$ we have still three broken
generators, but on the right hand side of the Nielsen--Chadha
inequality (\ref{NC}) we count $ 1 \times$ type I $+ 2 \times 2$
type II $=5$. This is larger than the number of broken generators,
which appears to be a novel observation. We note that five is
precisely the number of broken generators of the potential
symmetry group, which would correspond to the breaking pattern
$SO(6) \to SO(5)$.

We finally point out that the choice of coupling constants
discussed here gives rise to the ``polar" phase with a real order
parameter. In principle, an interaction with negative $\lambda'$
might give rise to ``ferromagnetic" type phases characterized by a
complex order parameter and a different symmetry breaking pattern
\cite{Volovik:2000mt}.

\section{Where to find vector condensates -- some examples}

We already mentioned the condensation of vector fields in
two--color QCD, which has been the main focus of our work. In
three color QCD at high quark chemical potential the 2SC phase
allows a spin-1 diquark condensate with albeit smaller gap than
the spin-0 condensate. This case has been recently investigated
\cite{2SC}. We also note that spin--1 condensation is expected to
give rise to a rich phenomenology of vortices
\cite{Volovik:2000mt}, which have been discussed in the context of
atomic Bose--Einstein condensates.

In the color--flavor locked phase of color superconducting quark
matter, the possiblilities of condensation of vector states in
presence of a nonvanishing isospin chemical potential are as yet
unexplored.

\section*{Acknowledgments}
The work of F.S. is supported by the Marie--Curie Fellowship under
contract MCFI--2001--00181. W.S. has been supported in part by
DAAD and NORDITA.

\end{document}